\documentclass[12pt]{iopart}

\expandafter\let\csname equation*\endcsname\relax
\expandafter\let\csname endequation*\endcsname\relax
\usepackage{amsmath}
\UseRawInputEncoding
\usepackage{graphicx}
\usepackage{subfigure}
\usepackage[ruled, linesnumbered]{algorithm2e}
\usepackage{setspace}
\usepackage{array}
\usepackage{booktabs}
\usepackage{tabularx}
\usepackage{multirow}
\usepackage{hyperref}
\usepackage{cleveref}
\usepackage{bm}
\hypersetup{hypertex=true,
	colorlinks=true,
	linkcolor=blue,
	anchorcolor=blue,
	citecolor=blue}
\usepackage{graphbox}

\usepackage{appendix}

\begin{document}

\title[]{First performance of hybrid spectra CT reconstruction: a general Spectrum-Model-Aided Reconstruction Technique (SMART)}

\author{Huiying Pan$^1$, Jianing Sun$^1$, Xu Jiang$^1$, and Xing Zhao$^{1,*}$}

\address{$^1$ School of Mathematical Sciences, Capital Normal University, Beijing, 100048, China}
\ead{* zhaoxing$\_$1999@126.com}

\vspace{10pt}
\begin{indented}
\item[Received] xxxxxx
\item[Accepted for publication] xxxxxx
\item[Published] xxxxxx
\end{indented}

\begin{abstract}
Hybrid spectral CT integrates energy integrating detectors (EID) and photon counting detectors (PCD) into a single system, combining the large field-of-view advantage of EID with the high energy and spatial resolution of PCD. This represents a new research direction in spectral CT imaging. However, the different imaging principles and inconsistent geometric paths of the two detectors make it difficult to reconstruct images using data from hybrid detectors. In addition, the quality reconstructed images considering spectrum is affected by the accuracy of spectral estimation and the scattered photons. In this work, firstly, we propose a general hybrid spectral reconstruction method that takes into account both the spectral CT imaging principles of the two different detectors and the influence of scattered photons in the forward process modelling. Furthermore, we also apply volume fraction constraints to the results reconstructed from the two detector data. By alternately solving the spectral estimation and the spectral image reconstruction by the ADMM method, the estimated spectra and the reconstructed images reinforce each other, thus improving the accuracy of the spectral estimation and the quality of the reconstructed images. The proposed method is the first to achieve hybrid spectral CT reconstruction for both detectors, allowing simultaneous recovery of spectrum and image reconstruction from hybrid spectral data containing scattering. In addition, the method is also applicable to spectral CT imaging using a single type of detector. We validated the effectiveness of the proposed method through numerical experiments and successfully performed the first hybrid spectral CT reconstruction experiment on our self-developed hybrid spectral CT system.

\vspace{1pc}
\noindent{\it Keywords}: multi-spectral computed tomography, basis material decomposition, iterative reconstruction, Schmidt orthogonal modification, nonlinear equations, inverse problem
\end{abstract}

\section{Introduction}
\label{introduction}

After decades of development, spectral CT has gradually become practical and is now widely used in different fields such as industrial non-destructive testing and medical diagnostics. Current spectral CT systems can be broadly divided into two categories based on the types of detectors used. The first category includes CT systems based on an energy integrating detector (EID), such as the Siemens SOMATOM Definition CT system and GE's Discovery CT750 HD CT system. The second category includes CT systems based on a Photon Counting Detector (PCD), such as Dectris' Pilatus, Eiger, Mythen PCD and the Medipix3 PCD developed by the European Organisation for Nuclear Research (CERN).

The imaging principles of EID and PCD are different. EID firstly converts X-rays into visible light, which is then converted into electric signals. This electric signals are integrated and finally converted into digital signals via A/D conversion for recording. Though EID uses an indirect detection method and do not have the ability to differentiate photon energy, they are larger in size and offer a wide scanning field of view. In addition, EID technology is well established and have lower equipment costs.When a single X-ray photon enters the PCD, it interacts with the semiconductor material to form electron-hole pairs. Under the influence of an external electric field, these are converted into electrical signals. The amplified electrical signals are then further amplified by the ASIC and compared to a predefined threshold value, after which they are recorded. PCD can distinguish the energy of individual photons, providing better energy resolution compared to the EID-based spectral CT systems. This improves material differentiation and quantitative accuracy. In addition, PCDs have smaller detector elements, resulting in higher spatial resolution in the reconstructed images. However, PCDs are typically smaller in size, which limits the scan field of view. Furthermore, due to current technological limitations, PCDs suffer from inconsistent detector element response, resulting in ring artefacts that negatively affect image quality in the reconstructed images.

A hybrid spectral CT equips both EID and PCD within a single CT system, offering the combined advantages of EID’s large scanning field of view and ring artifact-free reconstructed images, along with PCD’s superior energy and spatial resolution. This represents a new research trend in spectral CT imaging [11]. However, hybrid spectral imaging inevitably faces two key challenges: spectral estimation and the image reconstruction.

Existing spectral estimation methods can be divided into two categories: pre-estimation methods and iterative correction methods. In pre-estimation methods, a calibration phantom is scanned to estimate the spectrum in advance, which is then used for image reconstruction. The core of the pre-estimation methods is to map the different path lengths of the calibration phantom onto the corresponding projection data, thereby determining the mapping relationship. The iterative correction method alternately solves the spectral estimation problem and the image reconstruction problem, allowing both the spectrum and the reconstructed image to be recovered simultaneously. The Iterative approach have become a research hotspot in recent years.

Existing spectral CT reconstruction methods can be divided into two categories based on the solution domain: image-domain methods and projection-domain methods. This paper focuses on solving hybrid energy-spectrum CT imaging models. 

In addition to the challenges of spectral estimation and image reconstruction, hybrid spectral CT reconstruction faces several difficulties. First, the quality of the spectrally reconstructed image is affected by the accuracy of the spectral estimation and the influence of scattered photons. In addition, differences between detectors, such as imaging principles, detector element size and number, and scanning settings, make it difficult to reconstruct images using data from hybrid detectors. In response to the challenges of hybrid spectral CT reconstruction, the contributions of this paper are concluded as follows:
\begin{enumerate} 
\item[(\romannumeral1)]	A general energy-spectrum model-assisted hybrid energy-spectrum reconstruction method is proposed, which takes scattering into account in modelling the forward imaging process and allows simultaneous energy-spectrum recovery and image reconstruction from hybrid energy-spectrum data containing scattering;
\item[(\romannumeral2)]	The proposed method uses the ADMM method to solve the energy spectrum estimation and image reconstruction problems alternately, so that the energy spectrum and the reconstructed image can be corrected with each other, thus improving the accuracy of the energy spectrum estimation and the quality of the reconstructed image.
\item[(\romannumeral2)]	The proposed method takes into account the principle of dual-detector energy-spectrum CT imaging and is applicable not only to hybrid energy-spectrum CT imaging but also to single detector energy-spectrum CT imaging.
\item[(\romannumeral4)]	The proposed method realizes hybrid energy-spectrum CT reconstruction with two detectors for the first time, and the first hybrid energy-spectrum CT reconstruction experiment has been completed on our self-developed hybrid energy-spectrum CT system.
\end{enumerate}

\section{Method}
\label{method}

\subsection{Hybrid spectral CT model}
\label{hybrid spectral model}
The first step is to establish a cooperative optimization objective function, and imaging principles of different detectors, spectral estimation, scatter correction, and X-ray spectral CT reconstruction are all taken into account. In this section, details of the above part is described first, and then the objective function is given. Using $I$ and $C$ subscripts to represent the relevant variables of EID and PCD respectively.

\subsubsection{scatter-spectral model of EID}
\label{scatter-spectral model of EID}
X-ray intensity attenuates passing through the object, and part of  the reason for attenuation is scatter (within the medical energy range is mainly Compton scatter). Assume that scatter ratio to attenuation coefficient at $E_0$ keV is $\kappa(E_0)$, so the number of scatter photons received by the detector can be described as:
\begin{equation}
	\label{scatter number at detector}
	Sc=I_0\int s(E)\kappa(E)\int \mu(x,E)e^{-\int \mu(x,E)dx}dxdE.
\end{equation}

Considering the scatter photons, the scatter-spectral model of EID can be described as:
\begin{equation}
	\label{EID scatter-spectral model 1}
	\tilde{\phi}\left( \mu(x,E),s_I(E),\kappa_I(E),\sigma_I^2 \right)=-\ln\left( \int s_I(E)e^{-\int \mu(x,E)dx}dxdE +  (Sc_I/I_0)*g(x,\sigma_I^2)\right),
\end{equation}
where $g(x,\sigma_I^2)$ is a Gaussian kernel with variance $\sigma_I^2$ at $x$ position.

The X-ray spectral CT reconstrction of EID can be summarized as obtaining the spectrum $s_I(E)$, scatter ratio $\kappa_I(E)$ and  attenuation coefficient $\mu(x,E)$ of the scaned object from measured data $p_I$ by inversing the nonlinear model (\ref{EID scatter-spectral model 1}). However, (\ref{EID scatter-spectral model 1}) is difficult to solve causing the variables are coupled with each other, and needs to be simplified.

After the above steps, the simplified scatter-spectral model of EID can be expressed as:
\begin{align}
	\label{EID scatter-spectral model 2}
	\phi(f_m(x),s_I(E),\kappa_I)= -\ln \bigg( \int s_I(E)e^{-\sum_{m=1}^M\theta_m(E)\int f_m(x) dx}dxdE \nonumber \\
	+ \left(\kappa_Ie^{-p_I}\sum_{m=1}^M\theta_m(E)\int f_m(x) dx\right)*g(x,\sigma_I^2) \bigg) .
\end{align}

\subsubsection{scatter-spectral model of PCD}
\label{scatter-spectral model of PCD}
PCD response function is commonly represented by a Gaussian and polynomial cascaded model\cite{jilianPCD}:
\begin{align}
	\label{jilian}
	R(E;E',\omega(E'),\gamma(E')) &= \omega(E')R_G(E;E',\gamma(E')) + (1-\omega(E')) R_T(E;E') \\
	R_G(E;E',\gamma(E')) &= r_G(E;E',\gamma(E'))/\int r_G(E;E',\gamma(E'))dE \\
	R_T(E;E') &=  r_T(E;E')/\int r_G(E;E')dE \\
	r_G(E;E',\gamma(E')) &= \frac{1}{\sqrt{2\pi\sqrt{E'}\gamma(E')}}e^{-(E-E')^2/(2E'\gamma(E')^2)} \\
	r_T(E;E') &= \frac{1}{E'}\left[2(\frac{E-0.8E'}{(1-0.8)E'})^3-3(\frac{E-0.8E'}{(1-0.8)E'})^2+1\right],
\end{align}
where $\omega(E')$ and $\gamma(E')$ can be calibrated by radioisotope experiment.

The simplified scatter-spectral model of PCD can be expressed as:
\begin{align}
	\label{PCD scatter-spectral model}
	\psi(f_m(x),s_C(E),\kappa_C)=&-\ln \bigg( \int\int s_C(E)R(E;E')e^{-\sum_{m=1}^M\theta_m(E)\int f_m(x) dx}dxdE'dE \nonumber \\
	&+ \left(\kappa_Ce^{-p_C}\sum_{m=1}^M\theta_m(E)\int f_m(x) dx\right)*g(x,\sigma_C^2) \bigg).
\end{align}

\subsubsection{Volume fraction conservation}
\label{Volume fraction conservation}
On the one hand, due to the different imaging principles of EID and PCD, even with the setting such as the same X-ray source, the same voltage, and the same pre-filters, the received spectra from EID and different energy windows of PCD vary. Therefore, the equivalent attenuation coefficients of the received spectra are different, that means the reconstuction values are different. On the other hand, the scanning settings of the two detector systems, such as SOD, SDD, number of detector elements, size of detector elements, etc., are different, which make it almost impossible to constrain  data in the projection domain. Consequently, the volume fraction conservation constraint is considered in the image domain. Let $\rho_m$ represent the standard density of the $m$-th base material; the volume fraction of the $m$-th base material satisfies $f_m (x)⁄\rho_m \geq 0$, and the sum of the volume fractions of different base materials equals 1, i.e.,
\begin{equation}
	\label{volume conservation}
	V(f_m(x)) = \sum_{m=1}^M f_m(x)/\rho_m = 1.
\end{equation}

\subsubsection{Hybrid spectral CT model}
The known variates are measured data of EID and PCD, and the unknown variates are spectra $s_I(E)$, $s_{C_1}(E)$, $s_{C_2}(E)$, scatter ratio $\kappa_I$, $\kappa_{C_1}$, $\kappa_{C_2}$, and the basis material images $f_m(x)$. The general spectrum-model-aided cooperative optimization objective function is:
\begin{align}
	\label{final model}
	&(f_m(x),s_I(E),s_{C_1}(E),s_{C_2}(E),\kappa_I,\kappa_{C_1},\kappa_{C_2}) = \arg\min \frac{\alpha_1}{2} \left\| p_I - \phi \left( f_m(x),s_I(E),\kappa_I \right) \right\|^2 \nonumber \\
	& + \frac{\alpha_2}{2}\left\| p_{C_1} - \psi \left( f_m(x),s_{C_1}(E),\kappa_{C_1} \right) \right\|^2 + \frac{\alpha_3}{2} \left\| p_{C_2} - \psi \left( f_m(x),s_{C_2}(E),\kappa_{C_2} \right) \right\|^2 + \beta\cdot V\left( f_m(x) \right),
\end{align}
where $\alpha_t\in[0,1] (t=1,2,3)$ is the model coefficient, and $\sum_{t=1}^3 \alpha_t=1$. When $\alpha_1 = 0$, only PCD data is used; When $\alpha_2 = 0$ and $\alpha_3 = 0$, only EID data is used. When $\alpha_1\alpha_2 \neq 0$ or $\alpha_1\alpha_3 \neq 0$, mixed spectral data is used. $\beta$ is the parameter of the regular term.

\subsection{A general Spectrum-Model-Aided Reconstruction Technique (SMART)}
\label{section smart}
The ADMM method\cite{ADMM} was used to solve the optimization objective function (\ref{final model}) of hybrid spectral CT. Assume that after $k$ iterations, the estimated value is $(f_m^{(k)}(x),s_I^{(k)}(E),s_{C_1}v(E),s_{C_2}^{(k)}(E),\kappa_I^{(k)},\kappa_{C_1}^{(k)},\kappa_{C_2}^{(k)})$, and the solution of the $n+1$-th iteration can be obtained by solving the following two subproblems alternately:
\begin{align}
	\label{subprob1}
	\bm{Subproblem\,1}:& (s_I^{(k+1)}(E),s_{C_1}^{(k+1)}(E),s_{C_2}^{(k+1)}(E),\kappa_I^{(k+1)},\kappa_{C_1}^{(k+1)},\kappa_{C_2}^{(k+1)}) = \nonumber \\
	&\arg\min \frac{\alpha_1}{2} \left\| p_I - \phi \left( f_m^{(k)}(x),s_I(E),\kappa_I \right) \right\|^2 \nonumber \\
	&+ \frac{\alpha_2}{2}\left\| p_{C_1} - \psi \left( f_m^{(k)}(x),s_{C_1}(E),\kappa_{C_1} \right) \right\|^2 \nonumber \\
	& + \frac{\alpha_3}{2} \left\| p_{C_2} - \psi \left( f_m^{(k)}(x),s_{C_2}(E),\kappa_{C_2} \right) \right\|^2, \\
	\bm{Subproblem\,2}:& f_m^{(k+1)}(x) = \arg\min \frac{\alpha_1}{2} \left\| p_I - \phi \left( f_m(x),s_I^{(k+1)}(E),\kappa_I^{(k+1)} \right) \right\|^2 \nonumber \\
	& + \frac{\alpha_2}{2}\left\| p_{C_1} - \psi \left( f_m(x),s_{C_1}^{(k+1)}(E),\kappa_{C_1}^{(k+1)} \right) \right\|^2 \nonumber \\
	& + \frac{\alpha_3}{2} \left\| p_{C_2} - \psi \left( f_m(x),s_{C_2}^{(k+1)}(E),\kappa_{C_2}^{(k+1)} \right) \right\|^2 + \beta\cdot V\left( f_m(x) \right) .
\end{align}

\subsubsection{Solving subproblem 1}
\label{solve subprob1}
Apply the block coordinate descent (BCD) method\cite{BCD} to solve the subproblem 1, and the iteration scheme becomes:
\begin{equation}
	\label{solve s1}
	(s_I^{(k+1)}(E),\kappa_I^{(k+1)}) = \arg\min \frac{\alpha_1}{2} \left\| p_I - \phi \left( f_m^{(k)}(x),s_I(E),\kappa_I \right) \right\|^2,
\end{equation}
\begin{equation}
	\label{solve s2}
	(s_{C_1}^{(k+1)}(E),\kappa_{C_1}^{(k+1)}) = \arg\min \frac{\alpha_2}{2}\left\| p_{C_1} - \psi \left( f_m^{(k)}(x),s_{C_1}(E),\kappa_{C_1} \right) \right\|^2,
\end{equation}
\begin{equation}
	\label{solve s3}(s_{C_2}^{(k+1)}(E),\kappa_{C_2}^{(k+)}) = \arg\min \frac{\alpha_3}{2}\left\| p_{C_2} - \psi \left( f_m^{(k)}(x),s_{C_2}(E),\kappa_{C_2} \right) \right\|^2.
\end{equation}

Taking solving (\ref{solve s1}) as an example, the partial differentiation of (\ref{solve s1}) with respect to $s_I(E)$ and $\kappa_I$ respectively can be obtained:
\begin{align}
	&\frac{\partial}{\partial s_I(E)} = \alpha_1 \left( p_I - \phi \left( f_m^{(k)}(x),s_I(E),\kappa_I \right) \right) \frac{ e^{-\sum_{m=1}^M\theta_m(E)\int f_m(x)dx} } { \phi \left( f_m^{(k)}(x),s_I(E),\kappa_I \right) }, \\
	&\frac{\partial}{\partial \kappa_I} = \alpha_1 \left( p_I - \phi \left( f_m^{(k)}(x),s_I(E),\kappa_I \right) \right) \frac{ \left( e^{-p_I}\sum_{m=1}^M\theta_m(E)\int f_m(x)dx \right) * g(x,\sigma_I^2) } { \phi \left( f_m^{(k)}(x),s_I(E),\kappa_I \right) }.
\end{align}

The extreme point is obtained at the deviation of 0, that is, the following equations are solved:
\begin{equation}
	\label{solve s1 linear}
	\left( \frac{\partial}{\partial s_I(E)}, \frac{\partial}{\partial \kappa_I} \right)\left( s_I(E), \kappa_I\right)^\top = 0.
\end{equation}

Since the spectrum is non-negative and the EM method is sign-preserving, using EM to solve (\ref{solve s1 linear}).

\subsubsection{Solving subproblem 2}
\label{solve subprob2}
E-ART method\cite{EART} can obtain the basis material images using only one spectral data, and not need to register data. Therefore, the non-convexity projection-onto-convex-sets (NC-POCS) method\cite{ASDNCPOCS} based on E-ART is adopted to solve subproblem 2, and the iterative formula is as follows:
\begin{equation}
	f_m^{(k+\frac{1}{4})}(x) = f_m^{(k)}(x) + E-ART\left( p_I, f_m^{(k)}(x), s_I^{(k+1)}(E),\kappa_I^{(k+1)} \right),
\end{equation}
\begin{equation}
	f_m^{(k+\frac{2}{4})}(x) = f_m^{(k+\frac{1}{4})}(x) + E-ART\left( p_I, f_m^{(k+\frac{1}{4})}(x), s_{C_1}^{(k+1)}(E),\kappa_{C_1}^{(k+1)} \right),
\end{equation}
\begin{equation}
	f_m^{(k+\frac{3}{4})}(x) = f_m^{(k+\frac{2}{4})}(x) + E-ART\left( p_I, f_m^{(k+\frac{2}{4})}(x), s_{C_2}^{(k+1)}(E),\kappa_{C_2}^{(k+1)} \right),
\end{equation}
\begin{equation}
	f_m^{(k+1)}(x) = (1-\beta)f_m^{(k+\frac{3}{4})}(x) + \beta\left[ \rho_m\cdot (\sum_{m=1}^M f_m^{(k+\frac{3}{4})}(x)/\rho_m -1 )/M \right],
\end{equation}
\begin{align}
	E-ART\left( p, f_m(x), s(E),\kappa \right) &= [ \Theta_m ( p + \ln (\int s(E)\theta_m(E)e^{-\sum_{m=1}^M\theta_m(E)\int f_m(x)dx} \nonumber \\
	&+ (\kappa e^{-p_I}\sum_{m=1}^M\theta_m(E)\int f_m(x)dx ) * g(x,\sigma_I^2) )  ) ] / \left(\sum_{m=1}^M \Theta_m^2\right)
\end{align}
\begin{equation}
	\Theta_m = \frac{\int s(E)\theta_m(E)e^{-\sum_{m=1}^M\theta_m(E)\int f_m(x)dx}}{\int s(E)\theta_m(E)e^{-\sum_{m=1}^M\theta_m(E)\int f_m(x)dx} +  \left(\kappa e^{-p_I}\sum_{m=1}^M\theta_m(E)\int f_m(x)dx \right) * g(x,\sigma_I^2)}
\end{equation}

\subsubsection{Pseudo code of SMART}
\label{Pseudo}
\begin{algorithm}[htbp]
	\setstretch{1.1}
	\caption{Pseudo-code of the proposed SMART method}
	\label{alg1}
	\LinesNumbered
	\textbf{initialize:} measured $p_I$, $p_{C_1}$, $p_{C_2}$, $\sigma_I$, $\sigma_{C_1}$, $\sigma_{C_2}$, $\omega(E')$, $\gamma(E')$, threshold value of PCD bin $E_T$, initialized spectrum $s_0(E)$, $\alpha_1$, $\alpha_2$, $\alpha_3$, $f_m^{(0)}= 0$, $\kappa_I=\kappa_{C_1}=\kappa_{C_2}=0.001$, $s_I^{(0)}(E) = s_0(E)$, $\beta = 0.9$ \\
	$s_{C_1}^{(0)}(E)  = \int_{E_{min}}^{E_T} \int R(E;E',\omega(E'),\gamma(E'))s_0(E)dE'dE$ \\
	$s_{C_2}^{(0)}(E)  = \int_{E_T}^{E_{max}} \int R(E;E',\omega(E'),\gamma(E'))s_0(E)dE'dE$ \\
	\While{convergence conditions are not satisfied}{
		\For{$i=1$ to $I$}{
			 using extended-EM to solve (\ref{solve s1}) \\
			 using extended-EM to solve (\ref{solve s2}) \\
			 using extended-EM to solve (\ref{solve s3}) \\
		}
		$f_m^{(k+\frac{1}{4})}(x) = f_m^{(k)}(x) + E-ART\left( p_I, f_m^{(k)}(x), s_I^{(k+1)}(E),\kappa_I^{(k+1)} \right)$ \\
		$f_m^{(k+\frac{2}{4})}(x) = f_m^{(k+\frac{1}{4})}(x) + E-ART\left( p_I, f_m^{(k+\frac{1}{4})}(x), s_{C_1}^{(k+1)}(E),\kappa_{C_1}^{(k+1)} \right)$\\
		$f_m^{(k+\frac{3}{4})}(x) = f_m^{(k+\frac{2}{4})}(x) + E-ART\left( p_I, f_m^{(k+\frac{2}{4})}(x), s_{C_2}^{(k+1)}(E),\kappa_{C_2}^{(k+1)} \right)$\\
		$f_m^{(k+1)}(x) = (1-\beta)f_m^{(k+\frac{3}{4})}(x) + \beta\left[ \rho_m\cdot (\sum_{m=1}^M f_m^{(k+\frac{3}{4})}(x)/\rho_m -1 )/M \right]$\\
		$k=k+1$
	}
\end{algorithm}

\section{Experiment}
\label{experiment}
In this section, the influence of scatter data on spectrum estimation is studied by ablation experiment, and then the effectiveness of the proposed method is tested on simulated data. Finally, the first hybrid spectral CT reconstruction experiment is completed on the self-developed hybrid spectral CT system. All experiments were conducted on the same computer equipped with a 2.50GHz Intel i5-12400F six-core CPU and an NVIDIA GeForce RTX 3060 Ti graphics card.

\subsection{Spectrum estimation from scatter data}
\label{spectrum estimation scatter}
This section uses scatter data to test the effectiveness of the proposed method for spectrum estimation. The scan configuration is as follows: the distance between the X-ray source and the detector is 420 mm, the distance between the X-ray source and the rotation center is 300 mm, the number of detector units is 2068$\times$512, the size of detector units is 0.075$\times$0.075$\times$1 mm$^3$, the field of view is about 76.8 mm, the size of reconstructed image is 512$\times$512$\times$128, and the voxel size is 0.15$\times$0.15$\times$0.5 mm$^3$. The phantom is a bone phantom cylinder with a diameter of 4 cm and a height of 0.6 cm located at the center of the area to be reconstructed, with a standard density of 1.92 g/cm$^3$. The attenuation coefficient of the bone can be obtained from the website of the National Institute of Standard Technology (NIST).

\begin{figure}[htbp]
	\centering	
	\includegraphics[width=14cm]{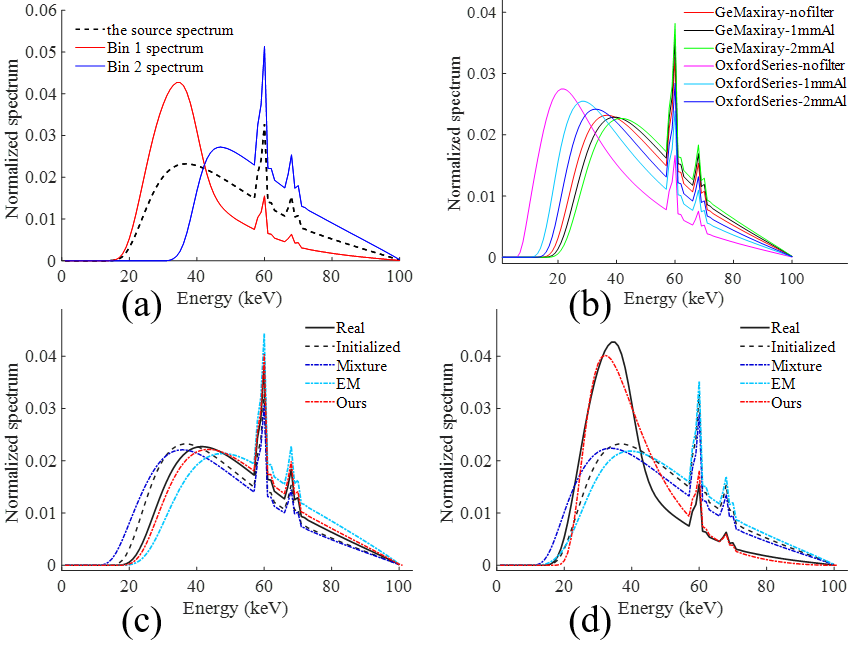} \hspace{-1cm}
	\vspace{-2.5mm}
	\caption{Initialized and estimated spectra in the ablation experiment. (a) Initialized spectra. (b)  Basis spectra of the Mixure method. (c) Estimated spectra of EID. (d) Estimated spectra of PCD.}
	\label{fig1}
\end{figure}

Obtain the source spectrum of GE Maxiray 125 at 100 kVp using Spectrum GUI, and simulate the reception spectrum of PCD at threshold [10,40] as the true value. Simulate scatter data using the open-source Monte Carlo code MC-GPU, and the initial number of photons is 40000 photons. Extract central layer data from scattered data and estimate the energy spectrum using different methods as shown in figure \ref{fig1}.

\begin{figure}[htbp]
	\centering	
	\includegraphics[width=14cm]{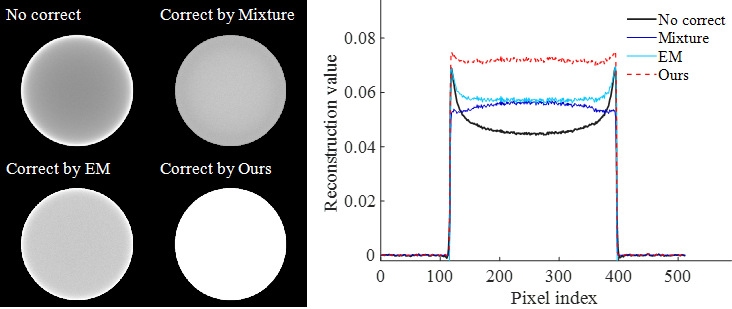} \hspace{-1cm}
	\vspace{-2.5mm}
	\caption{The results of the hardening correction using estimated EID spectra in figure \ref{fig1}. }
	\label{fig2}
\end{figure}

\begin{figure}[htbp]
	\centering	
	\includegraphics[width=14cm]{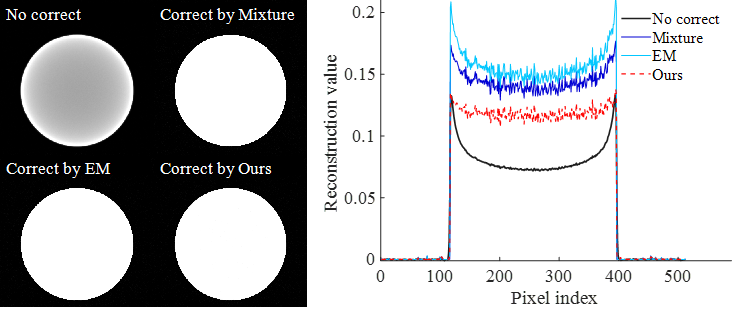} \hspace{-1cm}
	\vspace{-2.5mm}
	\caption{The results of the hardening correction using estimated PCD spectra in figure \ref{fig1}. }
	\label{fig3}
\end{figure}

Using estimated spectra in figure \ref{fig1} to perform hardening correction, the results are shown in figure \ref{fig2}-\ref{fig3}. On the left is the result of hardening correction, and the right is the 256th profile line.

\subsection{Real data experiment}
\label{Real data}

\begin{figure}[htbp]
	\centering	
	\includegraphics[width=14cm]{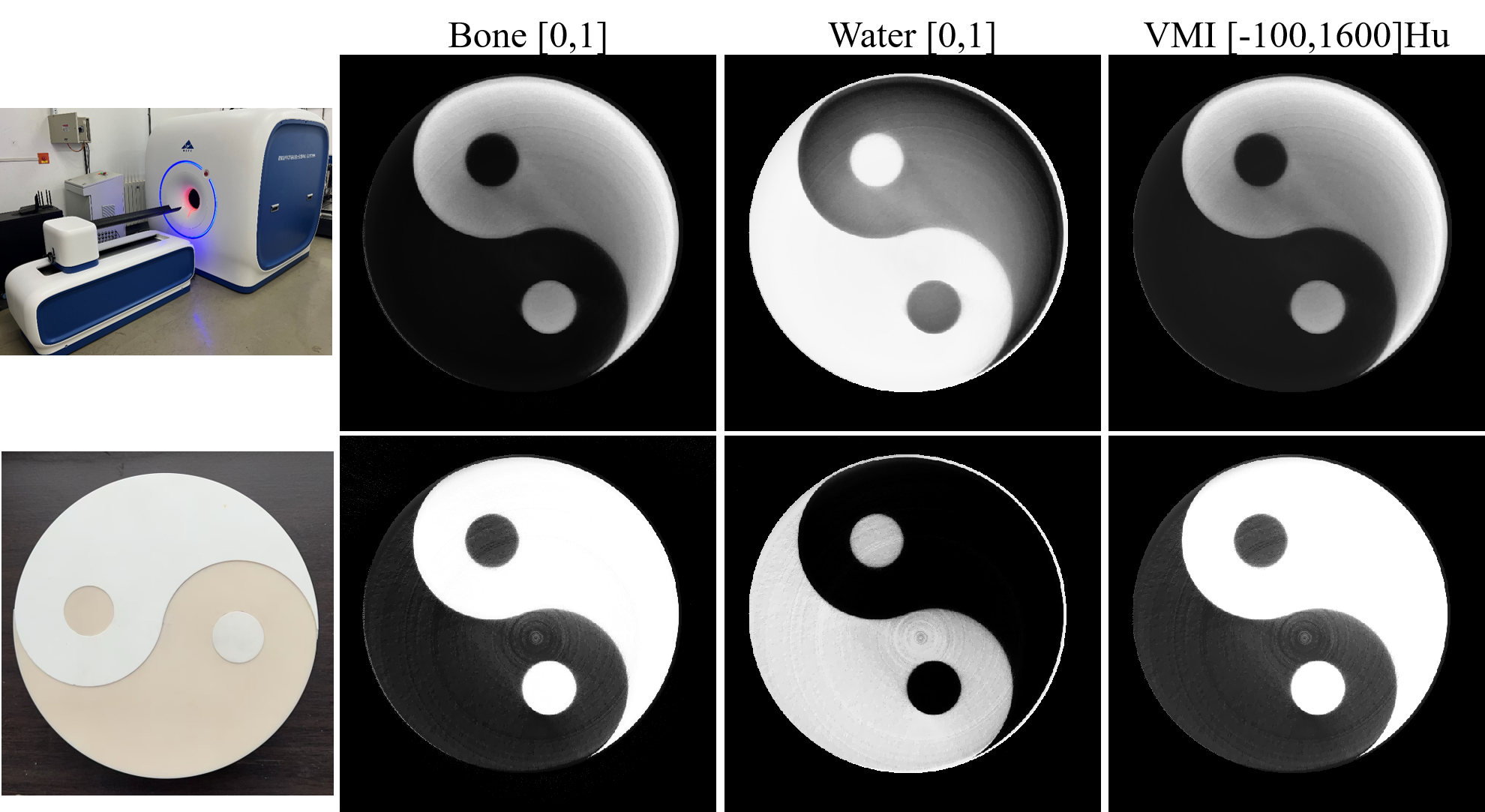} \hspace{-1cm}
	\vspace{-2.5mm}
	\caption{The self-developed hybrid spectral CT system and the results of hybrid spectral CT data. }
	\label{fig4}
\end{figure}

\section*{Reference}
%\References
\label{sect5}

\end{document}